# A Compact Online-Learning Spiking Neuromorphic Biosignal Processor

Chaoming Fang[1,2], *Student Member, IEEE*, Ziyang Shen[2,3], Fengshi Tian[2,4], Jie Yang[2], Mohamad Sawan[2], *Fellow, IEEE*

[1]Zhejiang University, Hangzhou, Zhejiang, China 310058

[2]CenBRAIN Lab., School of Engineering, Westlake University, Hangzhou, Zhejiang, China 310024

[3]School of Microeletronics, Fudan University, Shanghai, China 200433

[4]Department of Electronic and Computer Engineering, Hong Kong University of Science and Technology, Hong Kong, China

Email: yangjie@westlake.edu.cn, sawan@westlake.edu.cn

*Abstract*—Real-time biosignal processing on wearable devices has attracted worldwide attention for its potential in healthcare applications. However, the requirement of low-area, low-power and high adaptability to different patients challenge conventional algorithms and hardware platforms. In this design, a compact online learning neuromorphic hardware architecture with ultra-low power consumption designed explicitly for biosignal processing is proposed. A trace-based Spiking-Timing-Dependent-Plasticity (STDP) algorithm is applied to realize hardware-friendly online learning of a single-layer excitatory-inhibitory spiking neural network. Several techniques, including event-driven architecture and a fully optimized iterative computation approach, are adopted to minimize the hardware utilization and power consumption for the hardware implementation of online learning. Experiment results show that the proposed design reaches the accuracy of 87.36% and 83% for the Mixed National Institute of Standards and Technology database (MNIST) and ECG classification. The hardware architecture is implemented on a Zynq-7020 FPGA. Implementation results show that the Look-Up Table (LUT) and Flip Flops (FF) utilization reduced by 14.87 and 7.34 times, respectively, and the power consumption reduced by 21.69% compared to state of the art.

## I. INTRODUCTION

Reading and interpreting electrical biosignals like EEG [1], ECG [2], or EMG [3] has shown great potential in healthcare applications like monitoring, diagnosing, and disease treatment. With the rapid development of deep learning algorithms and big data trends, more and more robust and accurate performance is being achieved in interpreting biosignals [4]. Nevertheless, two main challenges exist when these deep learning algorithms are implemented on hardware. Firstly, low power consumption is preferred to conduct continuous monitoring of the biosignals, whereas deep learning algorithms are typically power-hungry and thus infeasible for such scenarios. Moreover, adaptability to different patients and conditions would greatly improve the robustness of the biosignal processing system, which requires on-chip learning to realize quick adaptation. However, the back-propagation (BP) learning algorithm for conventional neural networks takes up excessive energy and resources, which is inefficient for being implemented on wearable devices [5].

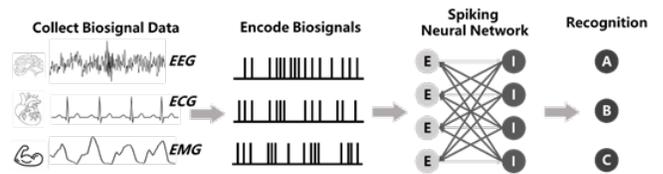

Fig. 1. Procedure for a typical neuromorphic biomedical signal processing.

Due to the intrinsic defects of conventional deep learning algorithms and their hardware implementation, neuromorphic computing has aroused global research interest in recent years. To train and inference a spiking neural network(SNN) is highly energy-efficient due to its high sparsity and local weight update mechanism [6, 7]. These features of neuromorphic computing make it an ideal choice for processing biosignals. As shown in Fig. 1, the raw biosignals are encoded into spike sequences and sent to SNNs to realize low-power recognition tasks. There have been many successful attempts to apply neuromorphic algorithms to biosignal processing. Amirshahi and Hashemi employed STDP and R-STDP algorithms to train the SNN model for ECG signal and achieved an accuracy of 97.9% [8]. Tian et al. utilized spiking CNN to predict the onset of Epileptic Seizure based on EEG data and reached a sensitivity of 95.1% while reducing 98.58% of computation complexity compared to CNN [9]. For the chip implementation of neuromorphic computing, there are also some representative works like Loihi [10], TrueNorth [11], SpiNNaker [12], ODIN [13], MorphIC [14]. However, these chips are mostly designed for large-scale simulation for neuroscience or other general-purpose computation tasks rather than specifically fabricated for biomedical application, and therefore occupy a relatively large area and power consumption, which is not applicable for the scenario of biosignal processing on wearable edge devices.

To fill the gap between the neuromorphic algorithms for biomedical applications and chip design. In this work, an online-learning neuromorphic biosignal processor with low power and utilization is proposed. An event-driven architecture and optimized computation logic are adopted to reduce power consumption and hardware resource overhead. A trace-based STDP rule is applied to realize hardware-friendly online learning. The proposed approach achieves a classification accuracy of 87.36% on the MNIST dataset and 83% on the ECG dataset. The hardware design is further verified on FPGA and reduced LUT, FF utilization, and power consumption by 14.87, 7.34 times, and 21.69% compared to existing works.

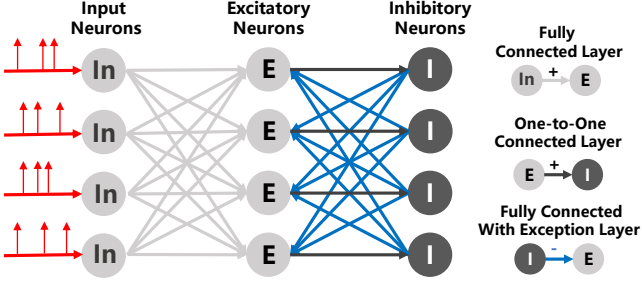

Fig. 2. Schematic of a single layer excitatory-inhibitory spiking neural network.

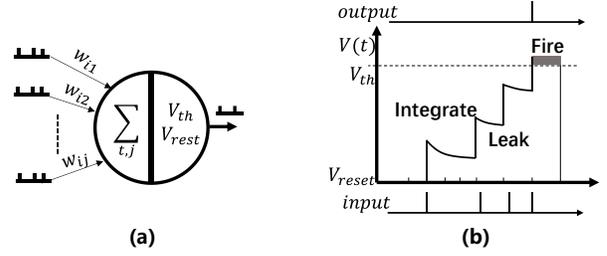

Fig. 3. Neuron model: (a) The computing scheme for a LIF neuron and, (b) The procedure of the voltage change of a LIF neuron.

The remaining parts of the paper are organized as follows. Section II describes the definition of the models of neuromorphic computing, including neurons, networks, and learning rules. Section III introduces the details of the proposed hardware design for the algorithm. Algorithm verification and FPGA implementation of the proposed design are demonstrated in Section IV.

## II. DESCRIPTION OF THE MODEL

### A. Spiking Neural Network

A typical SNN structure shown in Fig. 2 is made up of input, excitatory, and inhibitory layers [15]. Input neurons receive spikes and transmit them to excitatory neurons via a fully connected layer. Excitatory neurons integrate the spikes and fire another spike when the threshold potential is reached. Each excitatory neuron connects an inhibitory neuron with positive weights (grey arrows), and each inhibitory neuron, in turn, connects the entire neurons in excitatory layer with negative weights (blue arrows) except for the one previously connected to it. In this way, the lateral inhibition, also known as the winner-takes-all (WTA) mechanism, is achieved to ensure that neurons can learn to distinguish distinct features from the input spikes.

For the model of the neurons in SNN, the LIF model, also known as the Leaky-Integrate-and-Fire model, is a model with a good balance between hardware simplicity and biological plausibility. The LIF neuron in Fig. 3 (a) receives and generates spikes based on the membrane potential $V(t)$. The expression of $V(t)$ for LIF neurons is defined in [16], and its time dynamic behavior is depicted in Fig. 3 (b). At every time step, $V(t)$ is integrated by the sum of the input spike activation values and meanwhile decreased by a leaky term. When $V(t)$ is integrated into the threshold, a spike will be generated, and $V(t)$ drops to the pre-defined rest potential $V_{rest}$.

### B. Learning Algorithm

Spiking-Timing-Dependent-Plasticity (STDP) is one of the most popular unsupervised learning approaches for neuromorphic computing. The conventional STDP is defined as [17]:

$$\Delta w = \begin{cases} \sum_{t_{pre}} \sum_{t_{post}} A_{pre} e^{-(t_{post}-t_{pre})/\tau_{pre}}, & \text{if } \Delta t > 0 \\ -\sum_{t_{pre}} \sum_{t_{post}} A_{post} e^{(t_{post}-t_{pre})/\tau_{post}}, & \text{if } \Delta t < 0 \end{cases} \quad (1)$$

where $t$ denotes the spike timing of the neuron, $\tau$ is the decay constant, and $A$ is the amplitude constant. The subscript *pre* and *post* represent the pre and post-synaptic neurons. When conducting this conventional STDP learning approach, all the spike timings in pairs are required to be stored and accessed, which is not only inconvenient to be implemented in hardware design but also biologically unrealistic.

In this work, a variant of STDP known as the trace-based STDP is applied due to its efficiency in hardware implementation and is also proven to be equivalent to conventional STDP [18]. Specifically, an additional parameter *trace*, defined as $X(t)$, is added to capture the spike activity of the neuron. Its dynamics is given by:

$$\frac{dX(t)}{dt} = \frac{-X(t)}{\tau_x} + \alpha \sum_{i=1}^{m} \delta(t-t_i) \quad (2)$$

Each time a neuron generates a spike, its trace increases by $\alpha$, and otherwise, it decays by a factor of $\tau_x$ in the same exponential pattern as the $V(t)$. Here we define the trace of the input and excitatory neurons as $X_{pre}$ and $X_{post}$, respectively. The trace-based STDP rule is hereby defined. When the pre-neuron generates a spike, the weight $w$ is updated as:

$$w = w - \alpha_{post} X_{post}(t) \quad (3)$$

When the post-neuron generates a spike, $w$ is updated as:

$$w = w + \alpha_{pre} X_{pre}(t) \quad (4)$$

where $\alpha_{pre}$ and $\alpha_{post}$ denote the learning rates. We also define the reduction of $w$ as long-term depression (LTD), while the increase of $w$ as long-term potentiation (LTP). By conducting LTD and LTP at different spike events, the synapse weight can converge to stability after learning. Compared to the conventional STDP approach, trace-STDP learning only require access to the traces of a local pair of neuron for one synapse weight update, which minimizes memory access and computation loads. In addition, the weight update only occurs when a spike is generated, which is convenient for the implementation of an event-driven architecture.

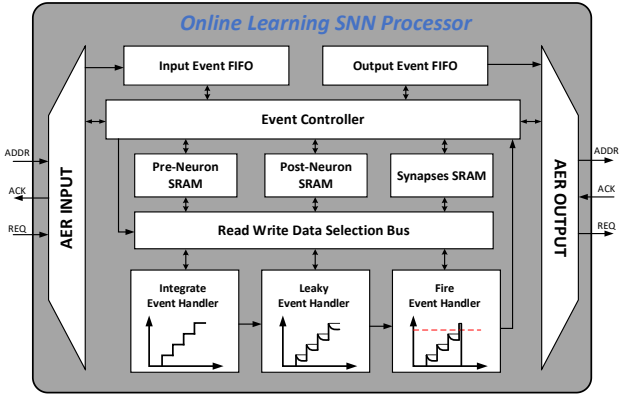
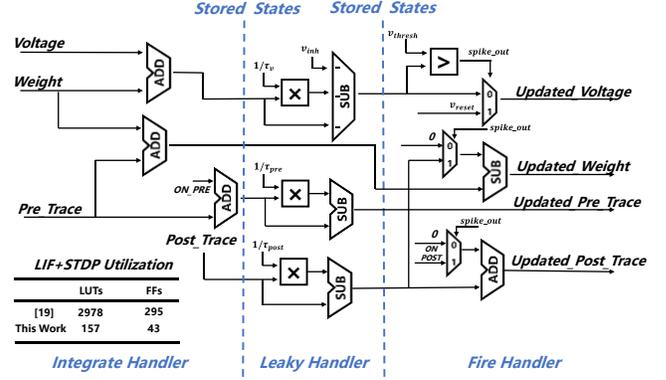

Fig. 4. Hardware architecture of the online learning spiking neural network processor.

Fig. 5. Handler circuits diagram and the comparison between the exponential and iterative decay circuits.

### III. PROPOSED HARDWARE ARCHITECTURE

#### A. Event-Driven Architecture

The proposed SNN model is implemented in an event-driven architecture shown in Fig. 4. Due to the requirement to minimize the chip area for biomedical applications, all the neurons and synapses' states are stored in SRAMs and updated in a time-multiplexing manner. Such a design approach is also suitable for low-power design due to the high sparsity of SNN. The input and output events are represented in an asynchronous protocol, Address Event Representation (AER), to handle real-time events. Every time an input spike is received or an output spike is generated, the AER bus generates an AER packet consisting of the spiking neuron ID and the corresponding timestamp. To realize real-time processing, two FIFOs are implemented as the buffer for the AER packets. Three event handlers work in an event-trigger mode and are only activated when the previous event is finished. For the rest of the time, the processor works in an idle state with low standby power. Such an event-driven architecture ensures the low-area and low-power features of the proposed SNN processor.

#### B. Controller and Event-Handler Design

The embedded controller is responsible for triggering different hardware resources to finish the algorithm proposed in Section II. The computation flow in the hardware implementation is given as follows. To start with, each AER packet consists of two parts, a timestamp and the neuron ID that generates a firing event. The controller consistently decodes the input AER packets from the input FIFO and judges if the input timestamp is equal to the current timestamp. If so, it means that the integration procedure for this timestamp has not been finished yet, and post-neuron voltage continues to integrate. Otherwise, it marks the end of the current timestamp so that fire and leak event handlers are triggered, and the neurons that satisfy the firing condition will be encoded into an output AER packets. Such an event-driven control scheme is able to process the incoming spike sequence generated from the analog frontend in real-time and dynamically adjust computation load according to input sparsity.

Three event handlers are implemented in the design, corresponding to the three stages of the LIF neuron's behavior. The controller activates the handlers in turn and accesses the memory data that the current handler needs. The handler updates the state information and stores it at the memory address where the original data is accessed. The detailed procedure is presented in Fig. 5. The integrate event handler decodes the input spike packet, accesses the synapse weights between the spiking neuron and post-neurons, and then integrates its weight into the $V(t)$ of the post-neurons. LTD update of the weights is also finished. After integration, the leakage event handler is activated to control the leak of the voltages and traces of all neurons. Additionally, the inhibition term is also subtracted from the voltage of excitatory neurons in this stage. Finally, the fire event handler judges which excitatory neuron would generate a spike by comparing their voltages to a pre-defined threshold. The LTP weight update is then triggered, and the spiking neuron ID and the current timestamp will be encoded as an output AER packet.

#### C. Neuron and Synapse State Update Circuit Design

The LIF neuron and trace-STDP synapse state update logic are given from Eq. (2) to Eq. (4). However, implementing these dynamic differential equations on hardware would consume excessive resources and power, which is not acceptable for compact and low-power designs. Therefore, the following discrete-time leaky equations are used to substitute the continuous ones in this work, which are given by:

$$X^* = X - \frac{dt \times X}{\tau_x} \quad (5)$$

$$V^* = V - \frac{dt \times (V - V_{rest})}{\tau_v} \quad (6)$$

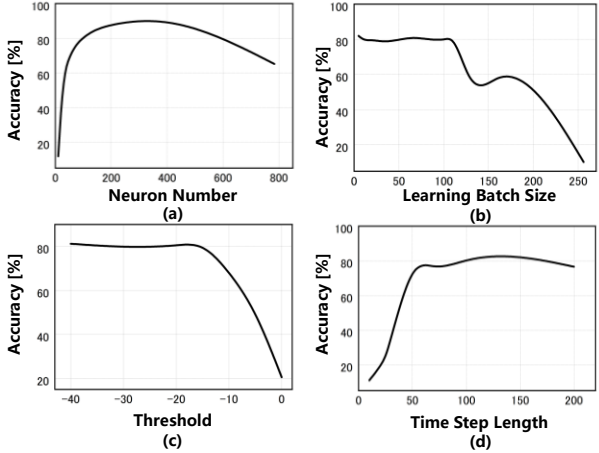

Fig. 6. Hyper-parameters discussion for the network: (a) Hidden neuron numbers, (b) Size of each learning batch, (c) Neuron voltage threshold, and (d) Length of timesteps.

The circuit implementation of Eq.(5) and Eq. (6) is given in Fig. 5. Such an implementation approach prevents too much use of complex hardware calculation resources like look-up tables for exponential calculation and can be easily implemented with a few fixed-point multipliers and adders. Compared to the work [19] implementing the same function with full precision, our implementation reduced the LUT and FF utilization by 18.97 and 6.87 times, respectively.

## IV. RESULTS AND DISCUSSION

To verify our proposed design and compare it with other works, we first apply our algorithm on the handwritten digits recognition dataset, which is typically used as the benchmark dataset for neuromorphic computing performance. Sixty thousand handwritten digits are encoded by Poisson distributed spike sequences and sent to SNN to do unsupervised online learning. Once the training is done, labels are assigned to the excitatory neurons based on the approach in [15], and another 10,000 digits are used for testing. Several hyperparameters of the SNN are discussed in Fig. 6 to achieve the best performance. Results show that the learning accuracy increases when the hidden neuron number and the timestep sequence length increase at first and then stop to increase and even go down a bit if the hyperparameters continue to grow. This is because the feature extraction ability of the network and encoder has reached saturation. For the voltage threshold, the accuracy drops when the threshold grows due to the fact that fewer firings will be generated by the excitatory neurons, and that would lead to failure in feature extraction. The influence on the accuracy of the change in learning batch size is similar to the change of the voltage threshold, mostly due to the under-fitting of the network at a large batch size. After carefully selecting the parameters according to this discussion, an accuracy of 87.36% has been reached.

We also evaluate the performance of our design on a four-class arrhythmia detection task using the MIT-BIH ECG dataset [20]. An SNN with the structure with 251 input neurons and 251 excitatory neurons is implemented, and the timestep of 100 is selected to fully encode the ECG raw signal.

TABLE I. COMPARISON OF THE FPGA IMPLEMENTATION RESULTS WITH EXISTING WORKS OF SPIKING NEUROMORPHIC HARDWARE

| Design | [13][a] | [19] | [21] | [22] | This Work |
|---|---|---|---|---|---|
| Learning Algorithm | SDSP | STDP | No | STDP | Trace-STDP |
| Accuracy | 84.5% | 89.1% | 93.8% | 90% | 87.36% (MNIST) 83%(ECG) |
| Clock Frequency (MHz) | 75 | 120 | 25 | 143 | 100 |
| FPGA Platform | Zynq-7020 | Virtex-6 | Spartan 6 | Kintex 7 | Zynq-7020 |
| LUT Utilization | 5114 | 71666 | 11489 | 5088 | **344** |
| FF Utilization | 4248 | 50921 | 4705 | 34646 | **579** |
| Neuron Number | 256 | 1591 | 1794 | 16 | 984 |
| Synapse Number | 65536 | 638208 | 647000 | 122 | 78400 |

[a.] The FPGA result is implemented using the open-source HDL codes at https://github.com/ChFrenkel/ODIN.git

Recognition accuracy achieves 83% after fine-tuning of the parameters, which demonstrates the potential of our design on edge biomedical signal monitoring and disease diagnosis.

Finally, we implemented our proposed hardware design on the Zynq 7020 FPGA. The implementation results and a comparison is summarized in Table I. Compared with the most compact design shown in [13], our work further reduced the LUT and FF utilization by 14.87 and 7.34 times, respectively, and dropped the power consumption by 21.69%.

## V. CONCLUSION

Neuromorphic computing has shown great potential in biosignal processing. In this work, a neuromorphic processor is proposed with an online learning function specifically designed for biosignal processing. A traced-based STDP algorithm is applied to realize hardware-friendly online learning. In the hardware domain, event-driven architecture and optimized computation logic are used to satisfy low power and utilization requirement. The result shows that the proposed design not only reaches the accuracy of 87.36% and 83% on the MNIST and ECG dataset but also manages to reduce LUT, FF hardware utilization, and power consumption by 14.87, 7.34 times, and 21.69%. The proposed design has great potential in applying to the wearable design for the healthcare application.

## ACKNOWLEDGEMENT

The authors would like to acknowledge the financial support and tools received from Westlake University and Zhejiang Key R&D Program No. 2021C03002 to support this project.